\documentclass[aps,preprintnumbers,prl,twocolumn]{revtex4}
\usepackage{amsfonts,amssymb,amsmath}            
\usepackage[]{graphics,graphicx,epsfig}            
\usepackage{amsthm}
\def\identity{\leavevmode\hbox{\small1\kern-3.8pt\normalsize1}}

\newcommand{\ket}[1]{\left | #1 \right\rangle}

\begin{document}

\title{Graph State Preparation and Cluster Computation with Global
  Addressing of Optical Lattices}
\date{\today}
\author{Alastair Kay}
\affiliation{Centre for Quantum
Computation,
             DAMTP,
             Centre for Mathematical Sciences,
             University of Cambridge,
             Wilberforce Road,
             Cambridge CB3 0WA, UK}
\author{Jiannis K. Pachos}
\affiliation{Centre for Quantum
Computation,
             DAMTP,
             Centre for Mathematical Sciences,
             University of Cambridge,
             Wilberforce Road,
             Cambridge CB3 0WA, UK}
\author{Charles S. Adams}
\affiliation{Department of Physics,
                        University of Durham,
			South Road,
			Durham DH1 3LE, UK}
             
\begin{abstract}
We present a novel way to manipulate ultra-cold atoms where four
atomic levels are trapped by appropriately tuned optical lattices.
When employed to perform quantum computation via global control, this
unique structure dramatically reduces the number of steps involved in
the control procedures, either for the standard, network, model, or
for one-way quantum computation. The use of a far-blue detuned
lattice and a magnetically insensitive computational basis makes the
scheme robust against decoherence. The present scheme is a promising
candidate for experimental implementation of quantum computation 
and for graph state preparation in one, two or three spatial dimensions.

\end{abstract}

\maketitle

The trapping and control of atoms within an optical lattice is
currently a topic of intense theoretical and experimental research,
with a view to simulating many-body systems \cite{Jaksch:a,Kukl} and
implementing quantum computation
\cite{Jaksch:c,brennen,Pachos:2003a}. While high fidelity
initialisation of suitable states has been
experimentally demonstrated \cite{Greiner:2003a, Peil}, there are
still significant barriers to a full scale implementation. Primarily,
the apparent requirement for individual addressing of atoms is a
major obstacle. To overcome this, a number of techniques are employed,
such as increasing the separation between lattice sites
\cite{scheun,Peil}, or using magnetic gradients \cite{khud}. Global
control \cite{Benjamin:2002a,Benjamin:2003b} is a conceptually
different approach that localises the action of global pulses to a
specific site of the lattice. Previous proposals require either long
initialisation times \cite{Vollbrecht} or employ superlattices
\cite{Zoller:1,Kay:2004a}. Another problem appearing in present
implementations is that two--qubit gates require the transportation
of entangled qubits around the lattice. This results in heating that
reduces the fidelity of the gate. The concept of a one--way
computation \cite{cluster1,cluster2} has been proposed to resolve this
problem. This involves performing appropriate measurements on any of a
special class of many-body entangled states, known as graph
states. The simplest example of such states, from the perspective of
experimental creation in optical lattices, is that of the cluster
state \cite{Greiner:2003a}.

In this letter, we present a realistic proposal for one--way quantum
computation in an optical lattice where the addressability problem is
resolved by global addressing. A significant reduction in the
complexity of the control sequences, and the principal novelty of the
scheme, is brought about by the appropriate tuning of optical
lattices. A far-off resonant lattice is used to trap four different
atomic states $\ket{0}$, $\ket{1}$, $\ket{P}$ and $\ket{P'}$. A
suitably tuned lattice can be added to move $\ket{P}$ and $\ket{P'}$
without perturbing $\ket{0}$ and $\ket{1}$. We could have chosen to
use these states to realise qudits ($d=3$ or $4$), and perform
computation directly with them. Instead, we keep the states $\ket{P}$
and $\ket{P'}$ in reserve to act as a unique, mobile, qubit which we
refer to as the pointer.
This allows us to construct very simple
control procedures that implement one-- and two--qubit gates,
measurements on individual qubits and create general graph states. The
simplicity of the
scheme allows us to employ more than one dimension to perform the
computation, thus significantly reducing the overall time needed to
transport information
through the lattice. Independent control of the pointer
position (with interferometric precision) facilitates the
implementation of both
one and two qubit gates without the need to ramp the magnetic field in
the vicinity of a Feshbach resonance \cite{Zoller:1}.

The computation is initialised by forming an optical lattice (labelled
$L$) with one atom in state $\ket{0}$ trapped in each lattice site
\cite{Greiner:2002a,Greiner:2003a}.
To initialise the pointer, we need to convert one of the atoms 
(or just add an extra one) into the $\ket{P}$ state. 
As an alternative to the techniques described in \cite{Vollbrecht,Kay:2004a}, we could transfer an
atom from the edge of the lattice to the state $\ket{P}$ and move it into the
computational region. To perform the transportation, we need an additional, state selective, lattice,
$L_P^+$ with a wavelength which is chosen such that the states
$\ket{0}$ and $\ket{1}$ are not perturbed.
Subsequently, the pointer is manipulated 
independently from atoms in the computational basis by $L_P^+$. 
The relative position of $L$ and $L_P^+$ is
adjusted with very high precision using interferometric techniques. We
can move the pointer around and cause it to interact with any single
qubit as we desire. The interaction between the pointer and the
target qubit results in phase accumulation that is conditional on the presence
of the pointer. In conjunction with Raman transitions applied to the
entire device, this interaction is sufficient to generate all the
control procedures we require.
\begin{figure}[!t]
\begin{center}
\includegraphics[width=0.35\textwidth]{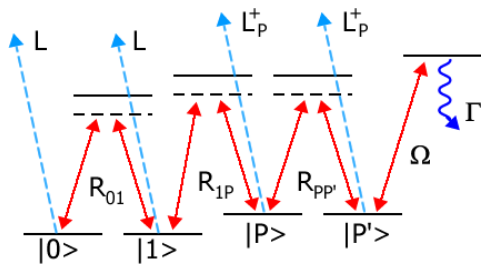}
\end{center}
\caption{Atomic level structure to be used in the computation.
The states of the computational basis, $\ket{0}$ and $\ket{1}$, are trapped by the lattice $L$ whereas $\ket{P}$ and $\ket{P'}$ are trapped by the laser $L_P^+$. The Raman transitions $R_{ab}$ act between the states $a$ and $b$.}
\label{lattice}
\end{figure}

\begin{figure}[!t]
\begin{center}
\includegraphics[width=0.38\textwidth]{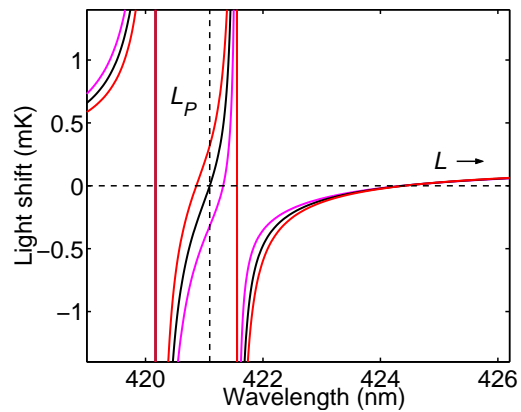}
\end{center}
\caption[light_shift]{The light shift in the vicinity of the $5s\rightarrow 6p$ transition 
as a function of the laser wavelength for the case of $^{87}$Rb.
The black curve corresponds to the qubit states, while
the other curves correspond to the pointer states
for left or right circularly polarised light.
The wavelengths of $L$ and $L_P$ are indicated.}
\label{fig:light_shift}
\end{figure}

To create the effective level scheme seen in Figure~\ref{lattice}, 
we trap alkali atoms (such as $^{87}$Rb) in an optical lattice, $L$, 
of wavelength $\lambda$ which is detuned by a few hundred nanometers from
the main atomic resonance. The logical basis of our qubits is encoded in the 
magnetically insensitive hyperfine sublevels of the ground state of 
the atom. For $^{87}$Rb, we select
$\ket{0}=\ket{F=1,m_F=0}$ and $\ket{1}=\ket{F=2,m_F=0}$.
The use of such states
greatly reduces the sensitivity of the computation to
external fields. 
For the pointer qubit, we select two of the magnetically
sensitive states (with the same magnetic quantum number).
In the far-detuned lattice, $L$, all the hyperfine ground states see the 
same trapping potential (see Fig.~\ref{fig:light_shift}). 
In order to move the pointer $\ket{P}$, we introduce an additional
lattice, $L_P$. This is composed of
a static circularly polarised component, $L_P^-$, that exactly
cancels $L$ and an orthogonally polarised component, $L_P^+$, 
whose phase controls the position of the
pointer. By tuning the frequency 
of $L_P$ within the fine structure splitting of the 
excited state, one can arrange that it does not perturb the qubit 
states. When using Rb, for example, we 
would select a wavelength of 421.1~nm, 
for the $5s\rightarrow6p$ transition (see Fig.~\ref{fig:light_shift}). 
In addition to the lattices $L$ and $L_P$, we will allow application of Raman transitions $R_{ab}$, between the states $a$ and $b$, to the whole lattice.

The central mechanism in our scheme requires an interaction between the
single pointer state, $\ket{P}$, and a particular, target, qubit.
Specifically, we are interested in creating a phase gate on the
target qubit, conditional on the presence of the pointer. There are
two different physical mechanisms that could be used, depending on
how the collisional energy shifts between the pointer and atoms
in states $\ket{0}$ and $\ket{1}$ compare ($U_0$ and $U_1$
respectively). In both cases, we move the pointer to the same lattice
site as the target qubit (by switching on $L_P$ and adjusting the
phase of $L_P^+$), induce the interaction, and then move the pointer
away again. If $U_0\neq U_1$, we can create a phase difference
$\int(U_1-U_0)dt=\pi$ ($\hbar=1$) simply by waiting for an appropriate time. 
The additional global phase of $\int U_0dt$ is irrelevant. Naturally,
this gate takes longer as the difference in energy decreases.
However, the difference between $U_0$ and $U_1$ can be made large by
working with an ambient magnetic field in the vicinity of an
appropriate interspecies Feshbach resonance. 

Alternatively, we can induce the interaction independently of the
values of $U_0$ and $U_1$ by driving a stimulated Raman transition,
$\tilde{R}_{1P}$, between the $\ket{1}$ state and a molecular bound state
\cite{Zoller:1,bound}. This is similar to inducing a Feshbach
resonance but, by using light, the speed of this
operation is only limited by the Rabi frequency of the free-to-bound
transition, which can be much higher than the collisional shift. A
$2\pi$ pulse flips the sign of the $\ket{1}$ component only at the
target site, where the resonant molecular bound state exists, thus
giving the required phase difference. In this scenario, the
collisional couplings ($U_0$ and $U_1$) are `always-on', and a
differential effective interaction, $U'_1\neq U_0'$, is induced by
light. 

Both techniques have the common property that they allow the
application of a phase gate to a single qubit by addressing the
entire structure (i.e. a localised phase gate). We will now
demonstrate how to combine this with Raman transitions, again
applied to the whole structure, to create all the elements required
for universal quantum computation.

The one--qubit gate, $U^\dagger\sigma_zU$, ($U \in SU(2)$) 
can be implemented on a target qubit in a straightforward way. The rotations 
$U$ and $U^\dagger$ are performed using the Raman transition $R_{01}$ 
on all the qubits. The pointer is used to create a localised
$\sigma_z$ operation, as has already been described.

The simplest way to perform a measurement on a specific qubit is to employ the 
fourth state, $\ket{P'}$.
Since this state is also transported by $L_P^+$, care has to be taken over 
the order of operations. Firstly, the pointer should be moved to the same 
lattice site as the target qubit,
then selective promotion of the target from  $\ket{1}$ to $\ket{P'}$
is achieved by performing a Hadamard in the $\{1,P'\}$ basis; 
a phase gate conditional on the presence of the pointer; and a second
Hadamard. A global measurement of the state $\ket{P'}$ can then 
be performed, providing information only about the state of the
target qubit. If found in this state, the qubit must be reset to the
$\ket{1}$ state. Finally, the pointer is moved away.

\begin{figure}[!t]
\begin{center}
\includegraphics[width=0.2\textwidth]{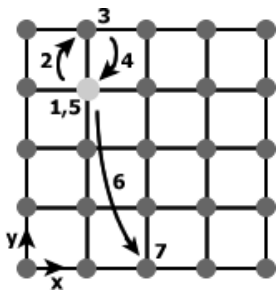}
\end{center}
\caption{Steps involved in a two--qubit gate. In Steps 1 and 5 the
  Hadamard rotation is applied with $R_{PP'}$. Steps 2, 4 and 6 are
  movements of the pointer (pale circle). Steps 3 and 7 are entangling
  procedures. After step 7, steps 1 to 6 must be repeated in reverse
  order to dis-entangle the pointer.}
\label{lattice2}
\end{figure}

A two--qubit gate also employs similar control procedures to those of
the one--qubit gate. In particular, we perform a Hadamard gate using
$R_{PP'}$, which
therefore only acts on the pointer. Suitable generation of phases by
collision with the control qubit, and repetition of the Hadamard with
$R_{PP'}$, causes selective deactivation of the pointer (by placing it
in $\ket{P'}$), dependent on the state of the control qubit. The
pointer qubit can then be moved to the target qubit to perform a
one--qubit gate (as seen in Fig. \ref{lattice2}), before undoing the
initial, entangling, steps. As a result, a general controlled-unitary
can be realised.

\begin{figure}[!t]
\begin{center}
\includegraphics[width=0.09\textwidth]{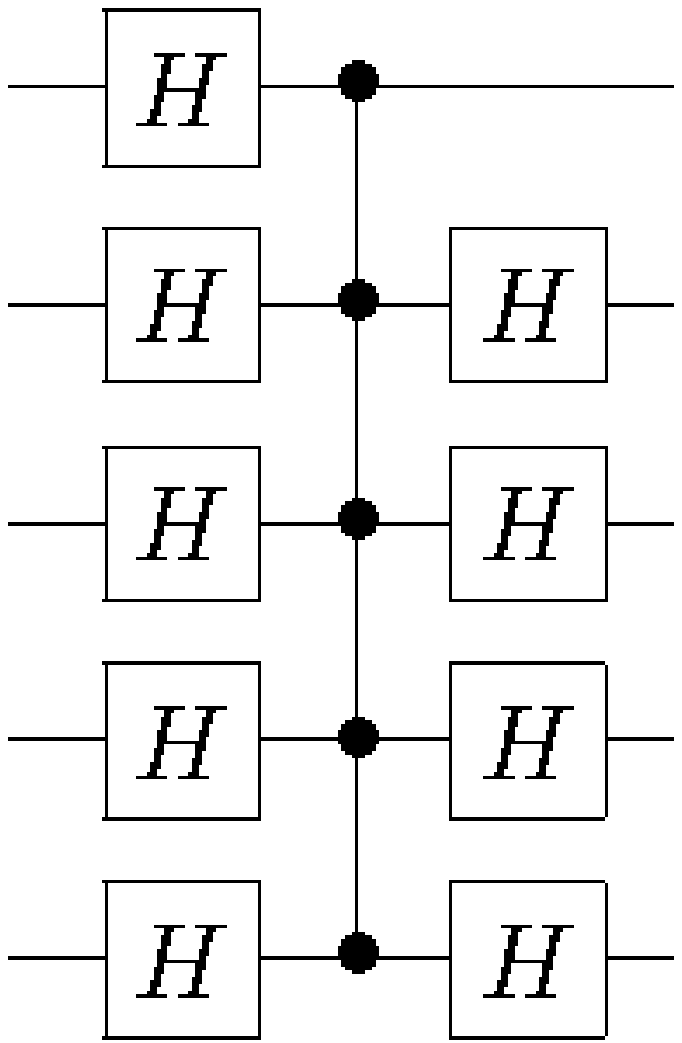}
\includegraphics[width=0.38\textwidth]{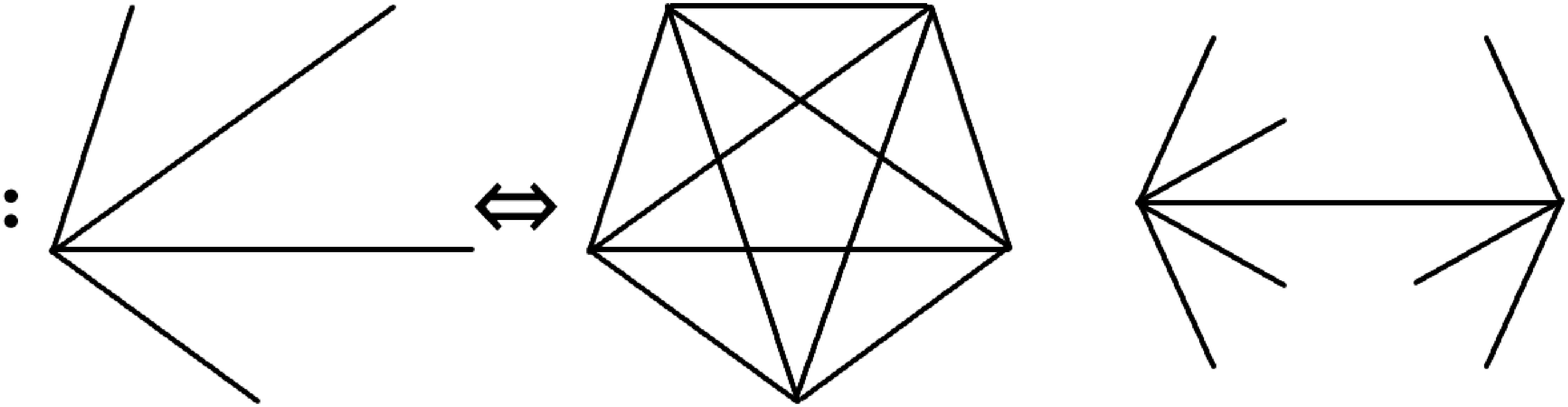}
\end{center}
\caption{The two first graphs represent the GHZ$_5$ state and are
  equivalent up to local
  unitaries. As the edges correspond to controlled-phase interactions, physical implementation of the
  first graph is simpler, requiring only one entangling step and a minimum of operations targeted by the entangled pointer. The third graph is an example that requires
  the pointer to be entangled a second time. }
\label{graph}
\end{figure}

This method for creating a two--qubit gate allows for a particularly efficient 
generalisation to multi-qubit gates where there is a single control and 
multiple targets. All that is required is to perform the entangling steps 1-5 
in Fig. \ref{lattice2} once, then repeat steps 6 and 7 for each target, 
before performing steps 5-1 to remove the pointer from the entangled state. 
A particularly simple example is to create graph states of the form $\ket{GHZ_n} 
= (\ket{0}^{\otimes n}+\ket{1}^{\otimes n})/\sqrt{2}$ between $n$ arbitrary 
qubits on the lattice.  As seen in Fig. \ref{graph}, this is achieved
by performing a global Hadamard, $H$, controlled phase gates on
$(n-1)$ of the qubits (all controlled by the $n^{th}$ qubit) and then
another global $H$. Finally, a local $H$ must be performed on the
$n^{th}$ qubit. At this stage
the pointer is disentangled and can be used to perform measurements on
the state. 
An arbitrary graph state can be prepared by applying this procedure an
appropriate number of times. In particular, we can take advantage of
the equivalence of the entanglement properties of graphs under local
unitaries (Fig. \ref{graph}) in order to minimise the number of edges,
each of which represents a controlled phase interaction \cite{hein}.

In order to perform a one-way quantum computation, we do not need to
perform any two--qubit gates. Instead, entanglement is used as an
initial resource by preparing the system in a cluster state. 
Here we can use $R_{01}$ to perform a Hadamard,
then $R_{1P}$ to promote the $\ket{1}$ components to 
$\ket{P}$ so that they can be transported, enabling collisions between neighbouring qubits. This involves
shifting the lattice independently by a single lattice site in the
$x$-and $y$-directions and
creating a phase of $\pi$ both times. Finally, we must reset the
$\ket{P}$ states to $\ket{1}$.

This completes the control set that is required, namely the ability to
perform single qubit rotations and measurements along with either
two--qubit gates or initialisation of a cluster state. All these
controls are simply built upon the ability to rotate the phase of one
of the trapping lattices in order to shift its minima, and to perform
Raman transitions and measurements on the whole lattice. The great
advantage in employing cluster states for computation stems from the
fact that the pointer, the one absolutely critical component in a
global control scenario, is never disturbed from its state
$\ket{P}$. It is not necessary to carry an entangled qubit around the
lattice, and thus the risk of decoherence is dramatically
reduced. Nevertheless, it is worth quantifying the additional
resources required. Consider, for example, performing a single qubit
rotation in a one--way computation. This is achieved by performing
single qubit rotations and measurements on 5 qubits (the single qubit
rotations are necessary because we have a fixed measurement
basis). Similarly, in order to perform a two--qubit gate between
distant qubits, the pointer has to be moved between the two. In the
case of a standard computation, the (entangled) pointer is moved over
approximately $2\sqrt{m}$ rows if the two qubits to be entangled are
separated by $m$ qubits on the circuit diagram (assuming computation
in 2D). In the case of the cluster computation, we have to move the
pointer over approximately $4m$ rows and $6$ columns, making $15m$
single qubit rotations and measurements \cite{cluster2}.

The threshold for fault-tolerant quantum computing places a stringent
demand on minimising decoherence. For optical lattices, far-blue
detuning \cite{blue} of the trapping light represents the most viable 
option: in this case atoms are
trapped at the minima of the light field, and decoherence due to
spontaneous scattering is reduced by an additional factor of order
$(2\pi a_0/\lambda)^2$, where $a_0$ is the harmonic oscillator length
of the trap ground state. For example, when considering Rb with a trap
frequency of 1~MHz, the spontaneous scattering rate for a lattice
wavelength between 428 and 590 nm is much less than $ 0.1~{\rm s}^{-1}$, 
and can be neglected compared to other sources of decoherence.

\begin{figure}[!t]
\begin{center}
\includegraphics[width=0.44\textwidth]{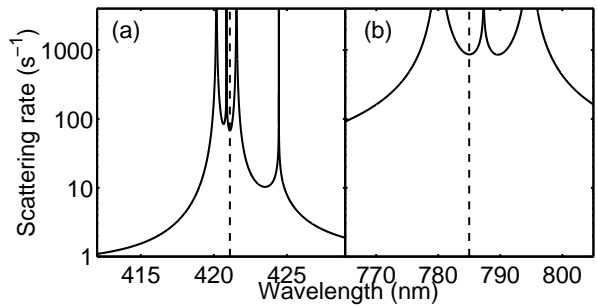}
\caption[scatter]{The spontaneous scattering rate at peak laser
intensity for an optical lattice with trap frequency of 1~MHz
in the vicinity of the (a) $6p$  and (b) $5p$ resonance 
of Rb. The wavelength of $L_P$ is indicated by a dashed line. 
The position of the narrow peak within the fine structure
splitting (corresponding to zero light shift) depends on the specific 
magnetic sub-levels and the laser
polarisation as shown in Fig.~\ref{fig:light_shift}.}
\label{fig:scatter}
\end{center}
\end{figure}

It is also necessary to ensure that the techniques employed to implement
computations do not introduce intolerable levels of scattering. 
For example, while moving the pointer, the qubit states are exposed to,
on average, half the intensity of $L_P^+$.
Fig.~\ref{fig:scatter} shows the scattering rate for a
fixed trap frequency of 1~MHz in the vicinity of the $6p$ and $5p$
resonances of Rb. The scattering rate is more than an order of
magnitude smaller for the weaker $6p$ resonance. For this reason we
choose the $5s\rightarrow 6p$ transition to manipulate the pointer
atom. The average scattering rate for our four states, due to the
state-selective lattice $L_P$ (at 421.1 nm) with
a trap frequency of 1~MHz, is less than $50~{\rm s}^{-1}$, so the probability
of spontaneous emission during a move time of 1~$\mu$s is less than
10$^{-4}$. If we relax the trap frequency by a factor of 10, we
decrease the computation speed by 10, but the spontaneous emission
rate is also reduced by the same factor, so the spontaneous emission
rate per gate operation is unchanged. Consequently, the optimum
trap frequency is a trade-off between decoherence due to spontaneous
emission and other mechanisms such as the excitation
of motional states. In our scheme, we reduce the sensitivity
to motional decoherence by keeping the computational qubits stationary throughout.

In \cite{Benjamin:2003b}, it was indicated that performing constant
measurements on some sections of a globally controlled device can
result in error suppression. The underlying mechanism is based on the
Zeno effect, which can help to stabilise classical states \cite{Kay:2004a}.
Unless we are
performing a two--qubit gate we can safely measure the $\ket{P'}$
state, forbidding transitions that might accidentally cause this state
to become populated. Such error suppression will
minimise the need for increasingly complex schemes such as error
correction.

In summary, we have proposed a system that allows the trapping and
manipulation of four atomic levels using two appropriately
tuned optical lattices. This structure enables the implementation
of a variety of quantum computation schemes, including a one--way
computation, that are based on global addressing, as well as the 
preparation of graph states between arbitrary qubits. The simplicity 
of the control procedures as well as significant suppression of some 
decoherence mechanisms render the present scheme as a plausible 
candidate for experimental realisation of quantum computation in 
optical lattices.

{\em Acknowledgements} J.K.P. would like to thank Hans Briegel for 
inspiring conversations and C.S.A. wishes to thank Simon Cornish for
stimulating discussions. This work was partially supported by EPSRC
and the Royal Society.

\end{document}